\documentclass[apjl]{emulateapj}
\usepackage{apjfonts}
\shorttitle{Spin of Spiral Galaxies} \shortauthors{Tonini et al.}
\journalinfo{Accepted by ApJL.}
\begin{document}
\title{Measuring the Spin of Spiral Galaxies}
\author{C. Tonini\altaffilmark{1}, A.
Lapi\altaffilmark{1,2}, F. Shankar\altaffilmark{1}, P.
Salucci\altaffilmark{1}}
\altaffiltext{1}{Astrophysics Sector, SISSA/ISAS, Via Beirut 2-4,
I-34014 Trieste, Italy} \altaffiltext{2}{Univ. ``Tor Vergata'', Via
Ricerca Scientifica 1, I-00133 Roma, Italy}
\begin{abstract}
We compute the angular momentum, the spin parameter and the related
distribution function for Dark Matter halos hosting a spiral galaxy.
We base on scaling laws, inferred from observations, that link the
properties of the galaxy to those of the host halo; we further
assume that the Dark Matter has the same total specific angular
momentum of the baryons. Our main results are: (i) we find that the
gas component of the disk significantly contributes to the total
angular momentum of the system; (ii) by adopting for the Dark Matter
the observationally supported Burkert profile, we compute the total
angular momentum of the disk and its correlation with the rotation
velocity; (iii) we find that the distribution function of the spin
parameter $\lambda$ peaks at a value of about $0.03$, consistent
with a no-major-merger scenario for the late evolution of spiral
galaxies.
\end{abstract}

\section{Introduction}

The mechanism of galaxy formation, as currently understood, involves
the cooling and condensation of baryons inside the gravitational
potential well provided by the Dark Matter (DM); in spirals, a
rotationally supported disk is formed, whose structure is governed
by angular momentum acquired through tidal interactions during the
precollapse phase.

Under the assumption of specific angular momentum conservation, that
holds when the baryons and the DM are initially well mixed, the
dynamics of the dark halo is directly related to the disk scale
length (see Fall \& Efstathiou 1980). This tight connection between
halo dynamics and disk geometry is quantified by the spin parameter
$\lambda$ (Peebles 1969).

The general procedure for the computation of the angular momentum
has been described in detail by Mo, Mao \& White (1998); it relies
on $3$ basic assumptions: (i) the mass of the galactic disk is a
universal fraction of the halo's; (ii) the total angular momentum of
the disk is also a fixed fraction of the halo's; (iii) the disk is
thin and centrifugally supported, with an exponential surface
density profile. The theory is applied to a Navarro, Frenk \& White
(1997; NFW) DM potential.

In the present work, we propose to determine the angular momentum
and the spin parameter of disk galaxies by making use of the
observed matter distribution in spirals, and of observed scaling
relations between halo and disk properties. For this purpose, we
adopt a modified set of assumptions: we relax (i), and use instead
an empirical relation that links the disk mass to that of its DM
halo (Shankar et al. 2005); we retain (ii) and suppose total
specific angular momentum conservation during the disk formation,
i.e., $J_D/M_D=J_H/M_{H}$ in terms of the disk and halo masses
$M_D$, $M_{H}$ and of the related total angular momenta $J_D$,
$J_H$; as to (iii), we still assume that the disk is centrifugally
supported, stable, and distributed according to an exponential
surface density profile, but we also take into account the gaseous
(HI+He) component. Finally, we perform the computation for a Burkert
halo.

We adopt a flat cosmology with matter density parameter
$\Omega_M\approx 0.27$ and Hubble constant
$h=H_0/100~\mathrm{km~s}^{-1}~\mathrm{Mpc}^{-1}=0.71$. Given a halo
with mass $M_{H}$ we determine its radius as $R_{H} = [3\, M_{H}\,
\Omega_M^z/$ $ 4\pi\,\rho_c\,\Omega_M\, (1+z)^3\,
\Delta_{H}]^{1/3}$; here $\rho_c=2.8\times 10^{11}\,
h^2\,M_{\odot}\, \mathrm{Mpc}^{-3}$ is the critical density,
$\Omega_M^{z}=\Omega_M\,(1+z)^3/[(1-\Omega_M)+\Omega_M\,(1+z)^3]$
and $\Delta_{H}=18\,\pi^2+82\,(\Omega_M^z-1)-39\,(\Omega_M^z-1)^2$
are the density parameter and the density contrast at redshift $z$;
$\Delta_{H}\approx 100$ holds at $z=0$.

\begin{figure*}
\plotone{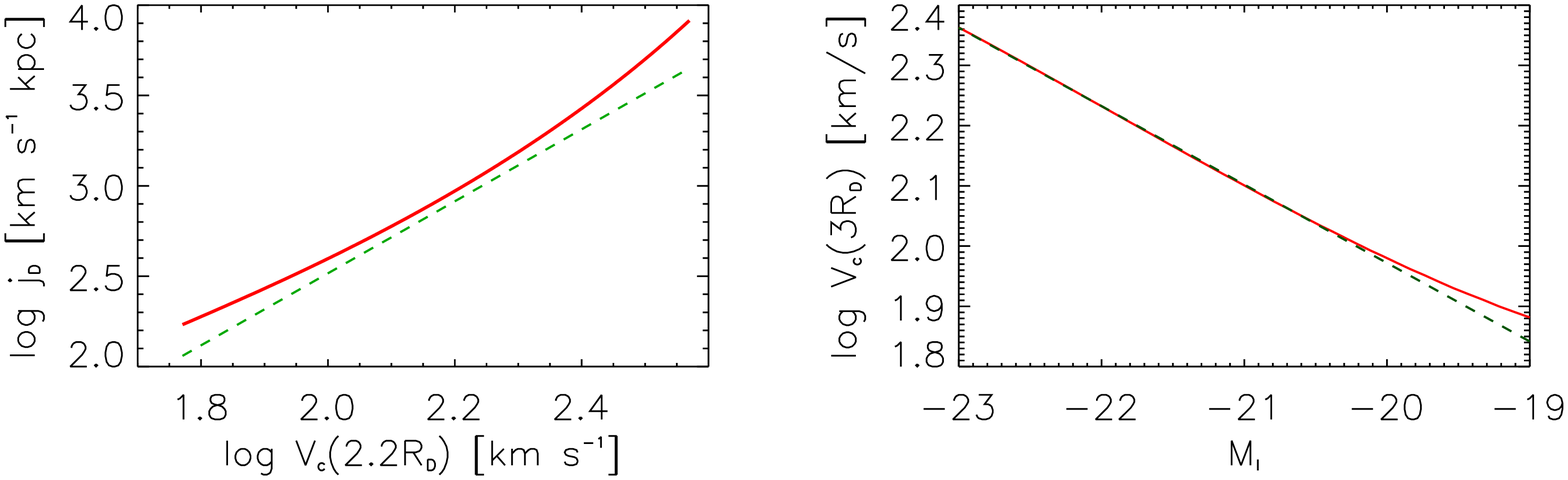} \caption{Left panel: the specific angular momentum
of the disk as a function of the rotation velocity at $2.2\,R_D$.
\textit{Solid} line is the result from this work, adopting the
Burkert profile; \textit{dashed} line is the best-fit relation from
the data collected by Navarro \& Steinmetz (2000), see their Figure
3. Right panel: the Tully-Fisher relation. \textit{Solid} line
represents the result from this work and \textit{dashed} line
illustrates the fit to the data by Giovanelli et al. (1997).}
\end{figure*}

\section{The angular momentum}

The fundamental parameters of the stellar disk and halo mass
distributions can be obtained straightforwardly by means of three
observational scaling relations, linking the disk mass to the halo
mass, to the halo central density, and to the disk scale length.

The total mass of the stellar disk $M_D$ that resides in a halo of
mass $M_{H}$ is given by the relation (Shankar et al. 2005):
\begin{equation}
M_{D} \approx 2.3 \times 10^{10} \ M_{\odot} \ \frac {(M_{H}/3 \
10^{11} \ M_{\odot} )^{3.1} }{1+(M_{H}/3 \ 10^{11} \
M_{\odot})^{2.2}}~; \label{MvirMd}
\end{equation}
this holds for halo masses between $10^{11}$ and about $3 \times
10^{12}\, M_{\odot}$, wide enough to include most of the spiral
population, except dwarfs. This relation has been derived by the
statistical comparison of the galactic halo mass function with the
stellar mass function; the related uncertainty is around $20\%$,
mainly due to the mass to light ratio used to derive the stellar
mass function from the galaxy luminosity function.

We model the stellar disk with a thin, exponential surface density
profile of the form
\begin{equation}
\Sigma_D(r) = {M_D\over 2 \pi \ R_D^2}\, e^{-r/R_D}~.
\end{equation}
The characteristic scale-length $R_D$ is estimated through:
\begin{equation}
\log{{R_D}\over \mathrm{kpc}} = 0.633+0.379\,\log{M_D\over
10^{11}\,M_{\odot}}+0.069\, \left(\log{M_D\over 10^{11}\,
M_{\odot}}\right)^2~; \label{mdrd}
\end{equation}
this relation is inferred from dynamical mass determinations by
Persic et al. (1996). These scale lengths are consistent with the
data by Dale et al. (1999), Simard et al. (1999), and Courteau et
al. (2003).

For the DM, we adopt a Burkert distribution $\rho_H(r)=\rho_0 \
R_0^3 / (r+R_0)(r^2+R_0^2)$, with $R_0$ the core radius and $\rho_0$
the effective core density. Correspondingly, the total halo mass
inside the radius $r$ is given by
\begin{equation}
M_{H}(<r)=4 \, M_0 \, \left[\ln\left( 1+\frac{r}{R_0} \right) -
\tan^{-1}\left(\frac{r}{R_0} \right) + \frac{1}{2}\,\ln\left(
1+\frac{r^2}{R_0^2}\right) \right]~, \label{mass}
\end{equation}
with $M_0=1.6\,\rho_0\,R_0^3$ being the mass contained inside the
radius $R_0$. The density $\rho_0$ is determined from the disk mass
through the relation obtained from the Universal Rotation Curve
(Burkert \& Salucci 2000):
\begin{equation}
\log{\rho_0\over \mathrm{g}~\mathrm{cm}^{-3}} = -23.515-0.964\,
\left({M_D}\over 10^{11}\,M_{\odot}\right)^{0.31}~. \label{rhoMd}
\end{equation}
For each given virial mass $M_H$ (corresponding to a radius $R_H$)
we find the density $\rho_0$ through Eqs.~(\ref{MvirMd}) and
(\ref{rhoMd}); then we numerically compute the core radius $R_0$ by
requiring that the mass $M_{H}(< R_{H})$ inside $R_H$ given by
r.h.s. of Eq.~(\ref{mass}) equals the virial mass $M_{H}$. The
resulting relation $R_0$ vs. $M_H$ is approximated within a few
percents by the relation
\begin{equation}
\log{(R_0/\mathrm{kpc})}\approx
0.66+0.58\,\log{(M_H/10^{11}\,M_{\odot})}~;\label{r0mh}
\end{equation}
such values of $R_0$ obtained by assuming a mass model out to $R_H$
are consistent with those estimated by Burkert \& Salucci (2000)
from the decomposition of the inner rotation curves.

The total circular velocity of the disk system is
$V^2_{c}(r)=V^2_D(r)+V^2_H(r)$. For a thin, centrifugally supported
disk the circular velocity is given by $V_D^2(r) = (G \, M_D/2\,
R_D)$ $\,x^2\,B(x/2)$; here $x=r/R_D$ and the quantity
$B=I_0\,K_0-I_1\,K_1$ is a combination of the modified Bessel
functions that accounts for the disk asphericity. Moreover, the halo
circular velocity is simply $V_H^2 (r)= G M_{H}(<r) / r$, and it is
useful to define $V_{H} = \sqrt{G M_{H}/R_{H}}$. Given the relations
(\ref{MvirMd}), (\ref{mdrd}), (\ref{rhoMd}) and (\ref{r0mh}) linking
the basic quantities of the system, the shape and amplitude of the
velocity profile depend only on the halo mass. Note that the
uncertainties on these relations combine to give a $10\%-20\%$
uncertainty on the determination of the velocity profile (see Tonini
\& Salucci 2004).

In order to check our mass model and empirical scaling relations, we
compute the I-band Tully-Fisher relation at $r=3\, R_D$. We obtain
the B-band luminosity from the stellar disk mass through the
relation $\log(L_B/ L_{\odot}) \approx 1.33+0.83 \, \log(M_D/
M_{\odot})$ by Shankar et al. (2005), then convert the related
magnitude in I-band through the mean colour $B-I\approx 2$ (Fukugita
et al. 1995). In Figure 1 (right) we compare the result with the
data by Giovanelli et al. (1997), finding an excellent agreement.

We compute the angular momentum of the disk as
\begin{equation}
J_D=2 \pi \int^{\infty}_0 \Sigma_D(r) \, r \, V_{c}(r)\, r\,dr = M_D
\, R_D \, V_{H} \, f_R\label{Jd}
\end{equation}
with $f_R = \int^{\infty}_0 x^2\, e^{-x} \, V_{c}(xR_D)/V_{H}~ dx$,
$x=r/R_D$ and $M_D=2 \pi \,\Sigma_0\, R_D^2$. Note that $J_D$
depends linearly both on the mass and on the radial extension of the
disk, while the DM distribution enters the computation through the
integrated velocity profile, encased into the shape factor $f_R$;
the latter slowly varies (by a factor $1.3$ at most) throughout our
range of halo masses.

In Figure 1 (left) we show the specific angular momentum of the disk
vs. the total circular velocity at $r=2.2\, R_D$, computed as
$j_D=J_D/M_D$ from Eq.~(\ref{Jd}). Plotted for comparison is also
the best-fit relation by Navarro \& Steinmetz (2000) from their
collection of data; note that these authors adopted a flat rotation
curve, so that $f_R=2$ and $j_D=2\,R_D\,V_H$.

We derive the halo angular momentum by assuming the conservation of
the total specific angular momentum between DM and baryons:
\begin{equation}
J_H=J_D \,\frac{M_{H}}{M_D}~, \label{haloangmom}
\end{equation}
an \textit{ansatz} widely supported/adopted in the literature (Mo et
al. 1998; van den Bosch et al. 2001, 2002; Burkert \& D' Onghia
2004; Peirani et al. 2004). Note that small variations of $J_D$ are
magnified by a factor $M_{H}/M_D$ in the value of $J_H$, i.e., the
latter is rather sensitive to the radial extension of the baryons.

We now consider, along with the stars, the gaseous component that
envelops the disk of spiral galaxies. We derive the total mass of
the gas component from the disk luminosity (see above) through the
relation
\begin{equation}
M_{\mathrm{gas}}= 2.13\times 10^6 \, M_{\odot}\,
\left(\frac{L_B}{10^6\, L_{\odot}}\right)^{0.81} \, \left[1-0.18 \,
\left(\frac{L_B}{10^8\, L_{\odot}}\right)^{-0.4}\right]~
\label{Mgas}
\end{equation}
by Persic \& Salucci (1999), where we have included a factor $1.33$
to account for the He abundance. Since the gas mass is on average
much less than the halo mass, both the total mass and the rotation
curve remain virtually unaltered ($V_{\mathrm{gas}} \sim
\sqrt{M_{\mathrm{gas}}/R_{\mathrm{gas}}}$) by the presence of the
gaseous component. However, the gas is much more diffuse than the
stars, reaching out several disk scale lengths (Corbelli \& Salucci
2000; Dame 1993); since most of the angular momentum comes from
material at large radial distances (van den Bosch et al. 2001), we
expect the gas to add a significant contribution to the total
angular momentum (Eq.~[\ref{Jd}]), especially in small spirals where
the gas to baryon fraction is close to $50\%$.

The detailed density profile of the gas in spirals is still under
debate in the literature. However, we are confident that the main
factors entering the computation of the gas angular momentum
$J_{\mathrm{gas}}$ are just the gas total mass $M_{\mathrm{gas}}$
and the radial extension of its distribution; in other words, we
expect that the details of the gas profile do not significantly
affect the results. In order to check this statement, we considered
$3$ different gas models, \textit{i.e.} (i) a disk-like distribution
(DL), with scale length $\alpha R_D$; (ii) a uniform distribution
(U) out to a radius $\beta R_D$; and (iii) an $M33$-like gaussian
distribution (M33; Corbelli \& Salucci 2000):
\begin{eqnarray}
\nonumber\Sigma_{\mathrm{gas}}^{\mathrm{DL}}(r) &=&
{M_{\mathrm{gas}}\over 2 \pi\, \alpha^2 R_D^2}\, e^{-r/\alpha
R_D}\\
\nonumber\\
\Sigma_{\mathrm{gas}}^{\mathrm{U}}(r) &=& {M_{\mathrm{gas}}\over
\pi\, \beta^2 R_D^2}\, \theta\left(r-\beta
R_D\right)\\
\nonumber\\
\nonumber\Sigma_{\mathrm{gas}}^{\mathrm{M33}}(r) &=&
{M_{\mathrm{gas}}\over \pi\, (2 k_1^2+k_2^2)\, R_D^2}\, e^{-(r/k_1
R_D)-(r/k_2 R_D)^2}~, \label{gas}
\end{eqnarray}
where $\theta$ in the second equation is the Heaviside step
function. As fiducial values of the parameters, we adopt
$\alpha\approx 3$ in the first expression, $\beta\approx 6$ in the
second one (Dame 1993), and $k_1\approx 11.9$, $k_2\approx 5.87$ in
the last one (Corbelli \& Salucci 2000). Each profile has been
normalized to the total gas mass $M_{\mathrm{gas}}$ as computed from
Eq.~(\ref{Mgas}).

In parallel with Eq.~(\ref{Jd}), the gas angular momentum will be
\begin{equation}
J_{\mathrm{gas}}= 2\pi\,\int_0^{\infty}\, \Sigma_{\mathrm{gas}}(r)\,
\, r \, V_{c}(r)\, r\, dr = M_{\mathrm{gas}} \, R_D \, V_{H} \,
f_{\mathrm{gas}}~,\label{Jgas}
\end{equation}
where the shape parameter $f_{\mathrm{gas}}$ encodes the specific
gas distribution. On comparing its values for the three models we
find differences of less than $15\%$, and so confidently choose the
gaussian profile as a baseline.

We then compute the halo angular momentum as a function of the total
baryonic one:
\begin{equation}
J_H=(J_D+J_{\mathrm{gas}}) \ \frac{M_{H}}{M_D+M_{\mathrm{gas}}}~.
\label{tothaloangmom}
\end{equation}
The gas is dynamically affecting the system mainly through its
different spatial distribution with respect to that of the stars,
adding an angular momentum component that is significant at large
radii compared to $R_D$.

Note that we do not include a bulge component, since it would
contribute a negligible angular momentum and a mass of $0.2\, M_D$
at most; in any case, this would slightly lower $J_H$ after
Eq.~(\ref{tothaloangmom}) and, as will be evident in the next
Section, would lower the spin parameter and strengthen our
conclusions.

\begin{figure*}[t]
\plotone{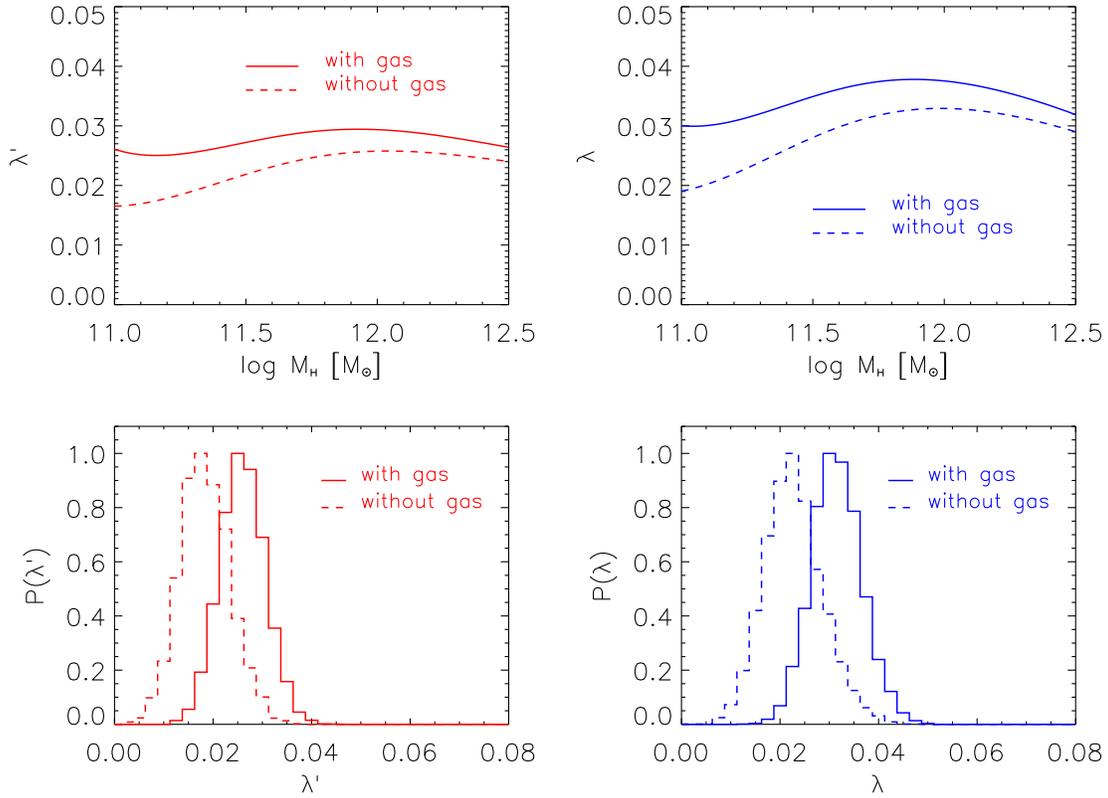} \caption{The spin parameter and its distribution
function. \textit{Top} panels: $\lambda'$ (\textit{left}) and
$\lambda$ (\textit{right}) as a function of the halo mass, when the
gas component is included in the system (\textit{solid} line) and
when it is not (\textit{dashed} line). \textit{Bottom} panels: the
distribution function of $\lambda'$ (\textit{left}) and $\lambda$
(\textit{right}), again with gas and without gas.}
\end{figure*}

\section{The spin parameter}

The spin parameter is a powerful tool to investigate galaxy
formation, as it is strictly related to both the dynamics and the
geometry of the system. We compute its values and distribution
function based on the results of Section 2.

The spin parameter is defined as follows:
\begin{equation}
\lambda=\frac{J_H\, |E_H|^{1/2}}{G \, M_{H}^{5/2}} \label{lambda}
\end{equation}
where $G$ is the gravitational constant, and $E_H$ is the total
energy of the halo. The latter is computed as
$|E_H|=2\pi\int{dr}\,r^2\,\rho_H(r)\,V^2_{c}$ after the virial
theorem, and having supposed that all the DM particles orbit on
circular tracks.

Bullock et al. (2001) proposed the alternative definition
\begin{equation}
\lambda'=\frac{J_{\mathrm{H}}}{\sqrt{2}\, M_{H} R_{H}\,
V_{H}}=\frac{J_H+J_D+J_{\mathrm{gas}}}{\sqrt{2}
(M_{H}+M_D+M_{\mathrm{gas}})\, R_{H} V_{H}}~, \label{lambdap}
\end{equation}
where the second equality holds after Eq.~(\ref{tothaloangmom}). We
find that for Burkert halos the ratio $\lambda/\lambda'$ is between
$1.1-1.3$ in the mass range $10^{11} - 3 \times 10^{12}\,
M_{\odot}$. In Figure 2 (top panels) we plot both $\lambda$ and
$\lambda'$ as a function of the halo mass. We highlight the
difference in the value of the spin parameter when the gas component
is included, especially in low mass halos.

To compute the probability distributions $\mathcal{P}(\lambda)$ and
$\mathcal{P}(\lambda')$ of the spin parameters, we make use of the
galactic halo mass function, i.e., the number density of halos with
mass $M_{H}$ containing a single baryonic core, as derived by
Shankar et al. (2005). A good fit is provided by the Schechter
function $\Psi(M_{H})=(\Psi_0/ \bar{M}) \, (M_{H}/
\bar{M})^{\alpha}\, \exp{(-M_{H}/\bar{M})}~,$ with parameters
$\alpha=-1.84$, $\bar{M}=1.12\times 10^{13}\, M_{\odot}$ and
$\Psi_0=3.1\times 10^{-4}\, $ Mpc$^{-3}$; note that within our range
of halo masses, this is mostly contributed by spirals. For the
computation of $\mathcal{P}(\lambda)$ or $\mathcal{P}(\lambda')$, we
randomly picked up a large sample of masses distributed according to
$\Psi(M_{H})$, then compute $\lambda$ or $\lambda'$ for each using
Eqs.~(\ref{lambda}) and (\ref{lambdap}), and eventually build up the
statistical distributions. During this procedure we have convolved
the relations (\ref{lambda}) and (\ref{lambdap}) with a gaussian
scatter of $0.15$ dex that takes into account the statistical
uncertainties in the empirical scaling laws we adopt; these are
mostly due to the determination of $R_D$ through Eq.~(\ref{mdrd}),
for which we have determined the scatter by using the disk mass
estimates of individual spirals reported in Persic \& Salucci
(1990).

As shown in Figure 2 (bottom panels), we find a distribution peaked
around a value of about $0.03$ for $\lambda$ and about $0.025$ for
$\lambda'$, when the gas is considered. We stress that this value of
$\lambda'$ is close to the result of the simulations by D'Onghia \&
Burkert (2004), who on average find $\lambda'=0.023$ for spirals
quietly evolving (i.e., experiencing no major mergers) since
$z\approx 3$, see their Figure 4. In addition, Burkert \& D'Onghia
(2005) argue that this value of $\lambda'$ provides a very good fit
to the observed relation between the disk scale length and the
maximum rotation velocity (see their Figure 1). We also stress that
our most probable value for $\lambda$ is in agreement with the
results by Gardner (2001), Vitvitska et al. (2002) and Peirani et
al. (2004), who find a peak value of the distribution function at
around $0.03$ for halos that evolved mainly through smooth
accretion.

\section{Discussion and Conclusions}

In this \textit{Letter} we have computed the angular momentum, the
spin parameter and the related distribution function for DM halos
hosting a spiral galaxy. We have relied on observed scaling
relations linking the properties of the baryons to those of their
host halos, and have assumed the same total specific angular
momentum for the DM and the baryons.

Our main findings are: (i) we show that including the gas component
beside the stars has a remarkable impact on the total angular
momentum; (ii) by adopting for the DM the observationally supported
Burkert profile, we compute the total angular momentum of the disk
and its relationship with the rotation velocity; (iii) we obtain
$\lambda'\approx 0.025$ and $\lambda\approx 0.03$ as most-probable
values of the spin parameters.

Simulations based on the $\Lambda$CDM framework, performed by
various authors (Bullock et al. 2001; D'Onghia \& Burkert 2004),
have shown that the distribution of the spin parameter $\lambda'$
for the whole halo catalogue peaks at around $0.035$, significantly
higher than our empirical value. However, D'Onghia \& Burkert (2004)
highlight that if one restricts one's attention to halos that hosts
spirals and have not experienced major mergers during the late
stages of their evolution ($z\la 3$), the average spin parameter
$\lambda'$ turns out to be around $0.023$, very close to our
observational result. Moreover, Gardner (2001) and Peirani et al.
(2004) showed that the spin parameter $\lambda$ undergoes different
evolutions in halos that have grown up mainly through major mergers
or smooth accretion: in the former case $\lambda$ takes on values
around $0.044$, while in the latter case $\lambda$ has lower values
around $0.03$.

Thus our findings point towards a scenario in which the late
evolution of spiral galaxies may be characterized by a relatively
poor history of major merging events.

\begin{acknowledgements}
We thank L. Danese, G. Gentile, and G.L. Granato for critical
reading, and the referee for her/his helpful comments. This work is
supported by ASI, INAF and MIUR.
\end{acknowledgements}

\end{document}